\documentclass[twocolumn,prl,showpacs,preprintnumbers,amsmath,amssymb,superscriptaddress]{revtex4}
\usepackage{graphicx}
\usepackage{color}
\usepackage{setspace}

\begin{document}

\title{Supersolid Phase Accompanied by a Quantum Critical Point in the Intermediate Coupling Regime of the Holstein Model }
\author{Yuta Murakami}
\affiliation{Department of Physics, University of Tokyo, Hongo, Tokyo 113-0033, Japan}
\author{Philipp Werner}
\affiliation{Department of Physics, University of Fribourg, 1700 Fribourg, Switzerland}
\author{Naoto Tsuji}
\affiliation{Department of Physics, University of Tokyo, Hongo, Tokyo 113-0033, Japan}
\author{Hideo Aoki}
\affiliation{Department of Physics, University of Tokyo, Hongo, Tokyo 113-0033, Japan}
\date{\today}

\begin{abstract}
We reveal that electron-phonon systems described by  
the Holstein model on a bipartite lattice exhibit, away from half filling, 
a supersolid (SS) phase characterized by coexisting charge order (CO) and superconductivity (SC), and an accompanying quantum critical point (QCP).  
The SS, demonstrated by the dynamical mean-field theory with a quantum Monte Carlo impurity solver, emerges 
in the intermediate-coupling regime, where the peak of the $T_c$ dome is located and the metal-insulator crossover occurs. 
On the other hand, in the weak- and strong-coupling regimes the CO-SC 
boundary is of first order with no intervening SS phases.  
The QCP is associated with the continuous transition from SS to SC and characterized by a reentrant behavior of the SS around it. We further show that the SS-SC transition is hallmarked by diverging charge fluctuations and a kink (peak) in the superfluid density.

\end{abstract}
\pacs{71.38.-k,71.10.Fd}

\maketitle

{\it Introduction}
---The competition between off-diagonal long-range order (ODLRO) and diagonal long-range order (DLRO)
is an issue of central interest in various classes of strongly correlated systems \cite{Gabovich2002, SS_boson,Anders2012,SS_spin,Matsuda2012,Ghiringhelli2012}. 
An important question is whether ODLRO and DLRO can coexist
and whether an associated quantum critical point (QCP) emerges.  
Specifically, the phase may be called a supersolid (SS) state when the DLRO 
is a charge order (CO) and the coexisting ODLRO  is superfluidity or superconductivity (SC). SS phases have been 
investigated in bosonic systems \cite{SS_boson}, typically liquid helium, 
but also in boson-fermion mixtures \cite{Anders2012} and spin systems \cite{SS_spin}. In electron systems, related phenomena have been observed. For example, it has been recently found that a compound in the iron-based superconductor 
family, $\text{BaFe}_2(\text{As}_{1-x}\text{P}_x)_2$, exhibits a QCP accompanied by a non-Fermi liquid behavior, which separates the SC state and a phase in which SC and antiferromagnetism (AF) coexist \cite{Matsuda2012}. For the cuprate superconductors, a possible existence of coexisting phases and/or a QCP is intensively discussed in relation to the pseudogap \cite{Ghiringhelli2012}.  

An important class of materials that provides an arena for competition between ODLRO and DLRO is a family of strongly interacting electron-phonon systems
with relatively large phonon energies and phonon-mediated attractive interactions 
comparable with the electronic bandwidth. 
A coexistence of $s$-wave SC and CO has been reported and discussed for Ba(Bi,Pb)O$_3$ and (Ba,K)BiO$_3$ \cite{Gabovich2002}. 
Further, the alkali-doped fullerenes \cite{c60} accommodate the competition between $s$-wave SC and AF. 
Although there are phenomenological arguments for 
explaining the coexistence of different phases \cite{Gabovich2002,McMillan1976}, a full understanding based on a
microscopic model is missing. 

In this Letter, we focus on the Holstein model on a bipartite lattice away from half filling to address the question of whether SS phases and QCPs exist in this simplest possible model for electron-phonon systems, 
without additional complexities (e.g., lattice frustrations or long-range interactions). 
There is in fact a long history of studies on strongly coupled electron-phonon systems based on the Holstein(-Hubbard) model \cite{Freericks1994,Freericks1993a,Tezuka2005,Scalettar1989,Noack1991,Alexandrov1986,Bauer2010,Murakami2013,Nowadnick2012,added1}.
The model is known to favor CO at half filling, 
while a SC phase emerges away from half filling 
\cite{Freericks1994,Freericks1993a,Tezuka2005,Scalettar1989,Noack1991,Alexandrov1986}.  However, the existence and stability of a SS phase in the Holstein
model have not been
established yet. Reference~\cite{Tezuka2005} studied the model in one dimension and showed that there is a coexisting region of SC and CO in the sense of a quasiordered phase in 1D. In Ref.~\cite{Alexandrov1986}, ordered states have been dealt with in the strong-coupling limit, but the possibility of phase separation has not been considered.

{\it Model and method} 
---We consider the Holstein model,
\begin{align}
 H = &-t\sum_{\langle i, j\rangle,\sigma}(c^{\dagger}_{i\sigma}c_{j\sigma}+{\rm H.c.})-\mu\sum_i(n_{i\uparrow}+n_{i\downarrow})\nonumber\\
 &+g\sum_i(b^{\dagger}_i+b_i)(n_{i\uparrow}+n_{i\downarrow}-1)+\omega_0\sum_ib^{\dagger}_ib_i,
  \label{eq:Hmodel}
\end{align}
where $i$, $j$ are site indices, $c_{i\sigma}^{\dagger}$ is the creation operator of an electron with spin $\sigma$, $b_i^{\dagger}$ is that of a phonon 
with frequency $\omega_0$, 
$t$ is the hopping parameter between nearest-neighbor 
sites, $n_{i\sigma}$ is the number of electrons, 
$\mu$ is the chemical potential,  and $g$ is the electron-phonon coupling.  The effective static phonon-mediated attractive interaction between electrons is $-\lambda\equiv -2g^2/\omega_0$. We note that this model becomes an attractive Hubbard model at $\omega_0\rightarrow \infty$, whose properties on bipartite lattices have been investigated in many contexts \cite{attractiveHubbard}. It has an SU(2) symmetry at half filling and can show SS behavior. However, the Holstein model lacks this symmetry for finite $\omega_0$, so that the existence of SS states is not {\it a priori} clear.

For a systematic investigation of the ordered phases in the model, we employ  
the dynamical mean-field theory (DMFT) \cite{Metzner,Georges,Georges_Kotliar} with a continuous-time quantum Monte Carlo (CT-QMC, hybridization 
expansion) method as an impurity solver \cite{Werner2006,Werner2007,Werner2011,Murakami2013}. With the CT-QMC method, we solve an impurity problem coupled to a superconducting bath and Einstein phonons \cite{Murakami2013}, using the Lang-Firsov transformation to exactly evaluate the phonon contribution \cite{Werner2007}.

Here, we focus on an intriguing regime where $\omega_0$ is comparable to the electronic 
bandwidth $W$.  
This situation is realized in carbon based compounds such as alkali-doped fullerenes or the recently found aromatic superconductors \cite{picene}.  
We employ a Bethe lattice with infinite coordination number, which has a semicircular density of states, 
$\rho_0(\epsilon)=(4/\pi W)\sqrt{1-(2\epsilon/W)^2}$ and use $W/4$ as the unit of energy. 
We consider $s$-wave SC and staggered CO as possible orders. 
The order parameters are, respectively, $\Phi_{\rm{SC}}=\frac{1}{N}\sum_{i}\langle c_{i\downarrow}c_{i\uparrow} \rangle$ and $\Phi_{\rm{CO}}=|n_{A}-n_{B}|/4$,
where $N$ is the total number of lattice sites, and $A$ and $B$ label sublattices.
The phase boundaries are identified by onsets of these order parameters. 
Green's functions are collected 
on a grid of $N_{\tau}=4\times 10^3$ points in the DMFT+CT-QMC calculations. We also note that there is no sign problem in our case.

\begin{figure}[t]  
     \centering
  \includegraphics[width=85mm]{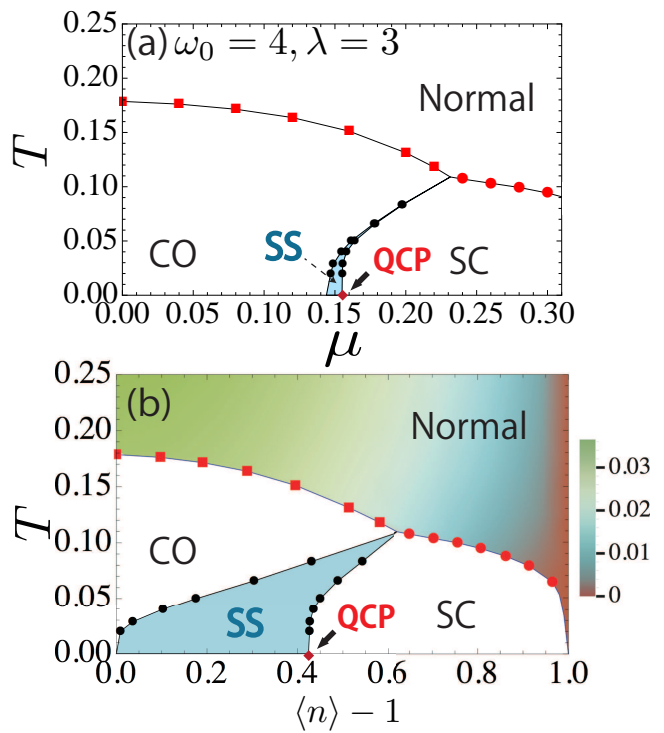}
 
  \caption{ (Color online) Phase diagram of the Holstein model  plotted against (a) the chemical 
potential $\mu$, and (b) the electron band filling $\langle n\rangle$ for $\omega_0=4, \lambda=3$. Blue areas indicate the supersolid (SS) region, 
while red diamonds at $T=0$ denote the quantum critical point (QCP). In the normal state,  the dc conductivity is displayed with color coding.  }
  \label{fig:phase_w4lam3}
  \end{figure}

{\it Results}
--- The main result of our Letter is the DMFT+CT-QMC phase diagram away from half filling, displayed in Fig.~\ref{fig:phase_w4lam3} for 
$\lambda=3$ and $\omega_0=4$. Fig.~\ref{fig:phase_w4lam3}(a) plots the phase boundaries against the chemical potential 
$\mu$, and Fig.~\ref{fig:phase_w4lam3}(b) against the electron band filling $\langle n\rangle$.

Notably, we find, in both panels, a SS region between the SC and CO phases in which the order parameters $\Phi_{\rm SC}$ (an ODLRO) and $\Phi_{\rm CO}$ (DLRO) are both nonzero. Since this SS phase appears in an extended region even on the $\mu$ axis, it should be robust against external fields and phase separation into SC and CO.  
The SS-SC phase boundary and SS-CO phase boundary are of second order as discussed below. 
 In both the $\mu$-$T$ and $n$-$T$ phase diagrams, the SS region widens as temperature decreases. For $T\rightarrow 0$, the SS phase appears at a nonzero value of $\mu$ ($\approx 0.145$), which corresponds to $\langle n\rangle=1$ (half filling), so that the SS phase appears 
immediately upon doping. 
The continuous transition between the SS and SC phases at finite temperatures suggests that this boundary ends at a QCP at $T=0$. We also note that the SS region is located below the CO phase [see Fig.~\ref{fig:phase_w4lam3}(a)(b)], and that in the filling range 
$0.43 < \langle n \rangle -1 < 0.6$, the SS emerges and then disappears as temperature is lowered (reentrant behavior). 
This behavior is qualitatively different from the 
phase diagram of $\text{BaFe}_2(\text{As}_{1-x}\text{P}_x)_2$
\cite{Matsuda2012}, where the SC+AF phase appears below both the SC and AF phases.
   
\begin{figure}[t]  
     \centering
   \includegraphics[width=85mm]{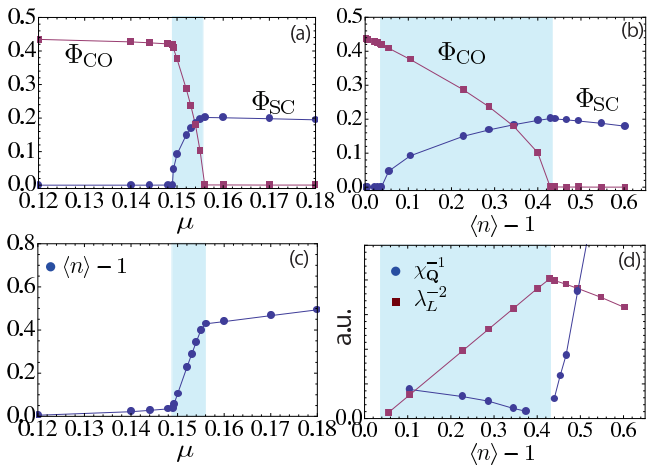}

  \caption{ (Color online) SC and CO  order parameters against $\mu$ (a) and against 
$\langle n\rangle$ (b).  Blue areas indicate the SS region. (c) Evolution of the filling as a function of $\mu$. 
(d) Inverse of the static charge susceptibility, $\chi_{\bf Q}$ at ${\bf Q}=(\pi,\pi)$, and the inverse of the squared London penetration depth $\lambda_L$, against the electron band filling. The parameters are $\omega_0=4, \lambda=3, \beta=35$. }
  \label{fig:u0lam3w4b35}
  \end{figure}

  To have a closer look at the behavior near the SC-CO boundary, we plot the order parameters in Fig.~\ref{fig:u0lam3w4b35} against $\mu$ [Fig.~\ref{fig:u0lam3w4b35}(a)] and 
against $\langle n \rangle$ [Fig.~\ref{fig:u0lam3w4b35}(b)], along with $\langle n \rangle$ 
vs $\mu$ [Fig.~\ref{fig:u0lam3w4b35}(c)] for $\lambda=3, \omega_0=4$, 
and inverse temperature $\beta=35$.  In the SS phase between the SC and CO phases ($0.149\alt\mu\alt0.156$), both $\Phi_{\rm SC}$ and $\Phi_{\rm CO}$ are indeed nonzero.  
Fig.~\ref{fig:u0lam3w4b35}(c) indicates that the compressibility $\partial n/\partial \mu$ is 
strongly enhanced in the SS.  More importantly, 
the SS and SC phases are not only continuously connected, but 
the static charge susceptibility $\chi_{\bf Q}$ 
at ${\bf Q}=(\pi,\pi)$, whose inverse is plotted 
in Fig.~\ref{fig:u0lam3w4b35}(d), diverges like $1/(\langle n\rangle-n_c)$ at the critical value $n_c$ for the SC-SS 
boundary. This divergence confirms the second-order nature of the phase transition. 
Here the susceptibility is computed by applying a small staggered external field $H_{\text{ext}}=\delta\mu (N_A-N_B)$, where $N_{A,B}=\sum_{i\in A,B} n_i$, with a tiny $\delta\mu=2\times10^{-4}$ for the SS and $\delta\mu=5\times10^{-4}$ for the SC phase. 
Another interesting quantity is the London penetration depth $\lambda_L$, 
because the superfluid density is proportional to $\lambda_L^{-2}=-(c^2/4\pi N)[\chi_{J,J}(i0^+)-e^2\sum_{{\bf k},\sigma}\langle (\partial^2 \epsilon ({\bf k})/\partial k^2_x) c^{\dagger}_{{\bf k},\sigma}c_{{\bf k},\sigma}\rangle]$. Here $\chi_{J,J}(\nu)$ is the current-current correlation function, $c$ is the speed of light, $e$ is the elementary charge and the lattice constant is set to unity. The general form of $\chi_{J,J}(i\nu_n)$ applicable to the SC, SS and CO phases is given in Ref. \cite{Supplemental}. The behavior of  $\lambda_L^{-2}$ in the SC and SS phases 
is shown in Fig.~\ref{fig:u0lam3w4b35}(d). We find that the SC-SS boundary 
is marked by a kink (maximum) in $\lambda_L^{-2}$, which remains when 
extrapolated to $T=0$ (see Ref.\cite{Supplemental}). 
In the SC region, the superfluid density increases towards half filling ($\langle n\rangle-1=0$),  
because the density of states of the free system increases near the Fermi energy, favoring SC, while 
in the SS region, the CO component grows towards half filling, weakening SC. 

At the CO-SS boundary, the order parameters also appear to be continuously connected, see Fig.~\ref{fig:u0lam3w4b35}. This suggests that the CO-SC boundary is of second order as well.
While the CO solution can be extended to larger fillings by suppressing SC, this solution is unstable against the introduction of a small SC component in the SS region. 
We also provide an argument why the free energy of the SS is lower than that for CO for $0.149\alt\mu\alt0.156$, see Ref. \cite{Supplemental}.


\begin{figure}[b]  
     \centering
   \includegraphics[width=85mm]{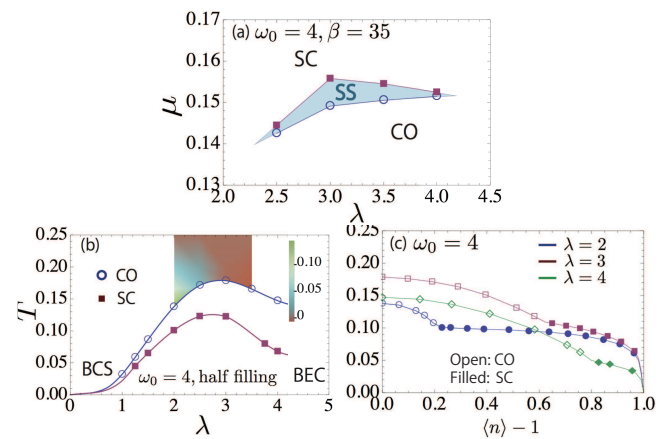}
  \caption{ (Color online) (a)  $\lambda$-$\mu$ phase diagram 
($\mu$: chemical potential, $\lambda$: phonon-mediated attraction) 
for the SS region for $\omega_0=4, \beta=35$. 
(b) Transition temperatures for CO and SC against $\lambda$ at 
half filling for $\omega_0=4$. CO is suppressed to obtain the $T_c$ for SC. In the normal state,  the dc conductivity (difference from its value at $T=0.25$ for each $\lambda$) is shown with color coding.
 (c) Transition temperatures for CO and SC as a function of filling for $\omega_0=4$ without any restriction on the type of order. }
  \label{fig:tc_w4}
  \end{figure}

In  Fig.~\ref{fig:tc_w4}(a)  we display the phase diagram in the space of $\mu$ and the phonon-induced attractive interaction $\lambda$  
for $\omega_0=4, \beta=35$. The SS region is widest around $\lambda=3$, while we find no SS phase for either
$\lambda=2$ or for $\lambda=4.5$, where a first-order transition (i.e., phase separation) occurs between SC and CO. 
The absence of the SS phase in the weak-coupling regime is consistent with perturbation theories \cite{Supplemental}\footnote{Let us note that the phase separation of the SC and CO phases in the weak-coupling regime is reminiscent of that between Fermi liquid and CO in the spinless Holstein model in the adiabatic limit. See, Ref.\cite{added1}.}.
We argue that the SS phase and associated QCP 
emerge only in the intermediate-coupling regime characterized by the peak of the $T_c$ dome
(which roughly corresponds to the BCS-BEC crossover region\footnote{We note that, strictly speaking, the peak of the $T_c$ dome does not necessarily coincide with the BCS-BEC crossover point, as is suggested for the attractive Hubbard model \cite{Toschi2005}.}).
In fact, if we look at the $\lambda$-dependence of the transition temperatures
for the SC and CO phases at half filling [Fig.~\ref{fig:tc_w4}(b)],
we find the peaks of the $T_c$ domes near $\lambda\simeq3$ for 
both the SC and CO phases at $\omega_0=4$. 
We also plot the transition temperatures as a function of the electron 
band filling in Fig.~\ref{fig:tc_w4}(c).
We can see that $\lambda=3$ has indeed the highest transition temperature at $\omega_0=4$, independent of filling.  Moreover, at this intermediate coupling ($\lambda\sim 3$),
a metal-insulator crossover occurs in the normal phase as one changes $\lambda$, where the insulator is a so-called bipolaronic insulator.
At $\lambda=3$, we observe that the dc conductivity [$\rm{Re} \sigma(0)$] increases with temperature, which is indicative of an insulating behavior, in the whole doping range [Fig.~\ref{fig:phase_w4lam3}(b)]. On the other hand, a metallic behavior appears around $\lambda=2.5$, which is shown for half filling in Fig.\ref{fig:tc_w4} (b) and in Ref. \cite{Supplemental} for the case away from half filling. We note that the SS region, the peak of the $T_c$ dome and the metal-insulator crossover point all shift in a correlated manner 
when $\omega_0$ is varied \cite{Supplemental}.

   \begin{figure}[h]  
     \centering
   \includegraphics[width=85mm]{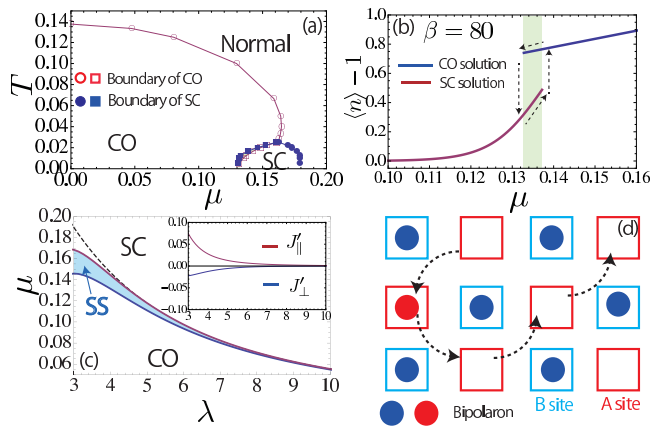}
  \caption{ (Color online) (a) Phase diagram of the mean-field solution for the leading-order effective spin model [Eq.~(\ref{eq:xxz})] at $\omega_0=4, \lambda=4.5$. (b) $\langle n \rangle -1$ as a function of $\mu$ at $\omega_0=4, \lambda=4.5, \beta=80$. The green area indicates a hysteretic region for the CO and SC solutions.  
(c) The phase diagram at $T=0$ of the 4th-order effective spin model. The blue region represents the supersolid state, and the dotted line the SC-CO boundary in  the leading-order effective spin model. The inset shows the dependence of the next-nearest neighbor exchange interaction on $\lambda$.  
(d) Schematic picture of the motion of bipolarons (circles) 
on the $A$ sublattice (red squares) with those on the $B$ sublattice (blue) forming a CO pattern.  Dotted arrows represent next-nearest neighbor 
hoppings arising from the higher-order terms in the effective spin model. }
  \label{fig:strong_coupling}
  \end{figure}

Physically, the emergence of the stable SS phase can be explained with the strong-coupling expansion, where long-range exchange interactions from high-order processes turn out to play a crucial role. To understand this, let us start from the lowest order $1/\lambda$ expansion \cite{Hirsch1983,Freericks1993}, which reduces
the Holstein model to an effective spin-$\frac{1}{2}$ $XXZ$ model with the nearest neighbor exchange interaction,
\begin{align}
&H_\text{eff}=-2\mu\sum_i S_i^{z}
-J_\perp\sum_{\langle i,j\rangle}(S_i^{x}S_j^{x}+S_i^{y}S_j^{y})+J_{\parallel}\sum_{\langle i,j\rangle}S_i^{z}S_j^{z}
, \label{eq:xxz}
\end{align}
where $J_\perp$ and $J_\parallel$ are functions of $g$ and $\omega_0$. 
$S^{+}\equiv S^{x}+iS^{y}$ and $S^z$ can be interpreted as a creation operator for a bipolaron and the corresponding density operator. In the 
limit of infinite spatial dimensions ($d\rightarrow \infty$), 
$J_{\perp}$ and $J_{\parallel}$ scale as $1/d$, and 
the mean-field solution of the effective spin model becomes exact. 
The result of the mean-field analysis is shown in Fig.~\ref{fig:strong_coupling}(a), where we have numerically solved the self-consistency equation.  
At $T=0$, the result is consistent with Ref.~\cite{Matsuda1970}: there is no finite SS region if one plots the phase diagram against $\mu$ (an external field in the spin model). Precisely at $\mu=(zJ_{\parallel}/4)\sqrt{1-(J_{\perp}/J_{\parallel})^2}$,  where $z$ is the coordination number, 
the SC, CO and SS phases become all degenerate. 
While this may seem to indicate a SS region 
if one plots the phase diagram against $n$ \cite{Alexandrov1986,Robaszkiewicz1981}, this occurs only at 
a single point on the $\mu$-axis, so that the SS is expected to be fragile against external perturbations and/or against phase separation to SC and CO.
At nonzero temperature, there is a finite 
hysteretic region where the solutions converge to either SC or CO [Fig.~\ref{fig:strong_coupling}(b)] with no intervening stable SS solutions.  
We thus conclude that, although the absence of SS is consistent with the QMC result in the strong coupling regime, the lowest-order $XXZ$ model \cite{Alexandrov1986,Robaszkiewicz1981} cannot explain the SS phase in the Holstein model without lattice frustration.

Now we consider the effect of higher-order terms. Generally, higher-order processes lead to longer-range exchange interactions as well as four- or more-spin interactions in the effective pseudospin model. While the details depend on the values of the parameters, it is known that longer-ranged exchange interactions can favor a SS \cite{Matsuda1970}. Indeed, the 4th-order expansion \cite{Freericks1993} on the Bethe lattice gives corrections of $J_{\perp}$ and $J_{\parallel}$ in Eq.~(\ref{eq:xxz}) and,
\begin{align}
H'_{\rm eff}=J_{\perp}'\sum_{\langle\!\langle i,j\rangle\!\rangle}(S_i^{x}S_j^{x}+S_i^{y}S_j^{y})+J_{\parallel}'\sum_{\langle\!\langle i,j\rangle\!\rangle}S_i^{z}S_j^{z},
\end{align}
where $\langle\!\langle\rangle\!\rangle$ denotes the next-nearest neighbor sites, and $J'_{\perp}$ and $J'_{\parallel}$ are negative and positive, respectively [inset of Fig.~\ref{fig:strong_coupling}(c)]. Intuitively, the $J_\perp'$ term enables bipolarons to move around on the $A$ sublattice while avoiding those on the $B$ sublattice 
forming a CO pattern, thus establishing a phase coherence within the less occupied sublattice in the CO background [Fig.~\ref{fig:strong_coupling}(d)]\footnote{The $B$ sublattice also has a SC component, and  the mean-field analysis suggests that it is stabilized by the $J_{\perp}$ term, which favors phase coherence between the $A$ and $B$ sublattices.}. In addition, within the mean-field theory it turns out that  the effect of the $J'_{\parallel}$ term can be translated to that of a $J'_\perp$ term with $J'_{\perp}=-J'_{\parallel}$ \cite{Matsuda1970}. Hence both terms cooperatively stabilize the SS phase in the intermediate-coupling regime,  where longer-range terms from higher-order processes become significant. As shown in Fig.~\ref{fig:strong_coupling}(c), the resultant mean-field phase diagram for the 4th-order effective pseudospin model indeed exhibits a SS region that widens toward the intermediate-coupling regime, although it overestimates the region.

   {\it Conclusion} 
   ---We have investigated the ordered phases in the Holstein model on a bipartite (i.e., nonfrustrated) lattice away from half filling 
with DMFT+CT-QMC calculation. 
We have focused on the unconventional region where $\lambda$ and $\omega_0$ are comparable to the bandwidth $W$.
Our study revealed that the intermediate coupling regime characterized by the peak of the $T_c$ dome and the metal-insulator crossover shows a supersolid phase and an associated QCP, 
while the observed absence of the SS in the weak- and strong-coupling regimes is consistent with perturbative analyses. 
 The continuous phase transition between the SC and SS phases is hallmarked by a downward  kink in the London penetration depth and a characteristic reentrance of the SS around the QCP.
We have also discussed that the stable SS phase is originated from long-range processes of bipolarons.
These results suggest that phenomena related to the SS phase and associated QCP may be explored in carbon-based compounds, some of which belong to the unconventional parameter regime considered here.
Further comparison with the related problem of AF+SC should help to understand the competition and coexistence of DLRO and ODLRO.

  {\it Acknowledgments.}
We thank D. Yamamoto and S. Yamazaki for helpful discussions.  Numerical calculations have been performed with a code based on ALPS~\cite{ALPS}. Y.M., N.T. and H.A. have been supported by
LEMSUPER (EU-Japan Superconductor Project) from JST, while PW acknowledges support from SNSF Grant 200021-140648 and FP7/ERC starting grant No. 278023.
Y.M. is supported by a Grant-in-Aid for JSPS Fellows.
\\

\section{\Large Supplementary Material }

\section{I. Conductivity and superfluid density}
Here we discuss the dc conductivity and superfluid density in the Holstein model. These quantities can be derived from the current-current correlation function,
for which we give 
the general expression 
applicable to normal, CO, SC and SS states. 
To investigate transport properties on the Bethe lattice, we adapt the formulas for the $d$-dimensional hypercubic lattice by substituting a semi-circular density of states \cite{Bethe_trans}. We can focus on the $x$-component $\chi_{J_x,J_x}$ without loss of generality. 

Within DMFT, the vertex correction vanishes due to the parity symmetry, and we can express the correlation function in terms of Green's functions as
\begin{align}
&\chi_{J_x,J_x}(i \nu_n)=\nonumber\\
&-\frac{e^2}{\beta}\sum_{ \omega_{n},\alpha,\alpha'}\int d\epsilon \Phi_{x,x}({\epsilon})\Big[G^{\alpha',\alpha}_{{\bf 0}}({\epsilon},i \omega_{n})G^{\alpha,\alpha'}_{{\bf 0}}({\epsilon},i \omega_{n}+i\nu_n)\nonumber\\
&- G^{\alpha',\alpha}_{{\bf Q}}({\epsilon},i \omega_{n})G^{\alpha,\alpha'}_{{\bf Q}}(-{\epsilon},i \omega_{n}+i\nu_n)\Big].
\end{align}
Here $\epsilon_{{\bf k}}$ is the energy of a free electron with momentum ${\bf k}$, $\Phi_{x,x}(\epsilon)\equiv\sum_{{\bf k}}(\partial \epsilon_{{\bf k}}/\partial k_x)^2\delta(\epsilon-\epsilon_{\bf k})$, $G_{{\bf q}}^{\alpha,\alpha'}(\epsilon_{\bf k},i \omega_{n}) = -\int_{0}^{\beta}\langle T_{\tau}c_{{\bf k},\alpha}(\tau)c^{\dagger}_{{\bf k+q},\alpha'}(0)\rangle e^{i\omega_n\tau}$ with $\alpha,\alpha'=\uparrow,\downarrow$.

For the Bethe lattice $\Phi_{x,x}(\epsilon) = (N/3d)[(W/2)^2-\epsilon^2]\rho_0(\epsilon)$ \cite{Bethe_trans}, where $N$ is the system size and $2d$ the coordination number. 
We note that the above expression at $\nu _n=0$ corresponds to $\chi_{J_x,J_x}(i 0^+)$ if the system is correlated.
 Since the dc conductivity in normal states is expressed as $\rm{Re}\,\sigma(0)=-\lim_{\nu\rightarrow 0^+}[\chi_{J_x,J_x}(i \nu)-\chi_{J_x,J_x}(i0^+)]/\nu$, we interpolate $\chi_{J_x,J_x}(i \nu_n)$ for $n=0,1,\cdots$ with polynomials of second-order, third-order or Pade approximations to evaluate $\rm{Re}\,\sigma(0)$.
In Fig.~1 of the main text, we show the results for the second-order interpolation, but all of these interpolations give qualitatively the same results. 

 The superfluid density is proportional to 
\begin{align}
\lambda_L^{-2}=&-(c^2/4\pi N)\Big[\chi_{J,J}(i0^+)\nonumber\\
&\hspace{10mm}-e^2\sum_{{\bf k},\sigma}\langle (\partial^2 \epsilon ({\bf k})/\partial k^2_x) c^{\dagger}_{{\bf k},\sigma}c_{{\bf k},\sigma}\rangle\Big],
\end{align} 
where $c$ is the speed of light, $e$ the elementary charge and the lattice constant is set unity. Here, we can make use of $\sum_{{\bf k}}(\partial^2 \epsilon_{\bf k}/\partial k^2_x)\delta(\epsilon-\epsilon_{\bf k}) = d \Phi_{x,x}(\epsilon)/d\epsilon$ to evaluate $e^2\sum_{{\bf k},\alpha}\langle (\partial^2 \epsilon ({\bf k})/\partial k^2_x)  c^{\dagger}_{{\bf k},\alpha}c_{{\bf k},\alpha}\rangle$ \cite{Bethe_trans}.

\subsection{A. Additional results for the conductivity}
\begin{figure}[h]  
     \begin{center}
  \includegraphics[width=80mm]{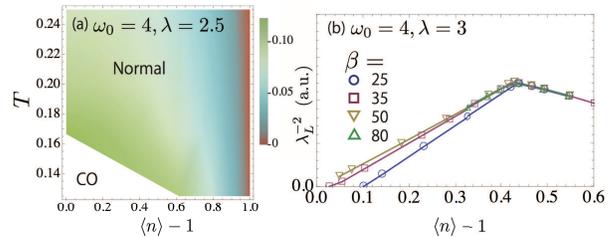}
  \end{center}
  \caption{(Color online) (a) Color-coded dc conductivity in the normal state for $\omega_0=4, \lambda=2.5$.  (b) Temperature dependence of $\lambda_L^{-2}$ for $\omega_0=4, \lambda=3$ around the SC-SS boundary. }
  \label{fig:fig_cond}
  \end{figure}

In Fig.~\ref{fig:fig_cond} (a), we show the conductivity in the normal state at $\omega_0=4, \lambda=2.5$ and away from half filling. This plot, together with Fig.~1(b) of the main text ($\omega_0=4, \lambda=3$),
 demonstrates that the metal-insulator transition at $\omega_0=4$ occurs between $\lambda=2.5$ and $\lambda=3$, independent of doping. 
Independent of filling, the conductivity for $\lambda=2.5$ decreases as the temperature is increased, which suggests that the normal state is metallic here.
The opposite behavior was found for $\lambda=3$ (see main text).

\subsection{B. Temperature dependence of the superfluid density}
Figure~\ref{fig:fig_cond}(b) illustrates the temperature dependence of the superfluid density in the SC and SS phases. 
As pointed out in the main text, there is a kink (maximum) in $\lambda_L^{-2}$ at the SC-SS boundary for all the temperatures investigated.
Since there is no significant temperature dependence near the transition point at temperatures lower than $T=1/\beta=1/50$ (Fig.~\ref{fig:fig_cond}), the  kink is a characteristic feature of the low-temperature regime, including the $T=0$ QCP. We also note that the 
temperature dependence of the superfluid density is much weaker in the SC phase than in the SS.

   \begin{figure}[h]  
     \centering
  \includegraphics[width=80mm]{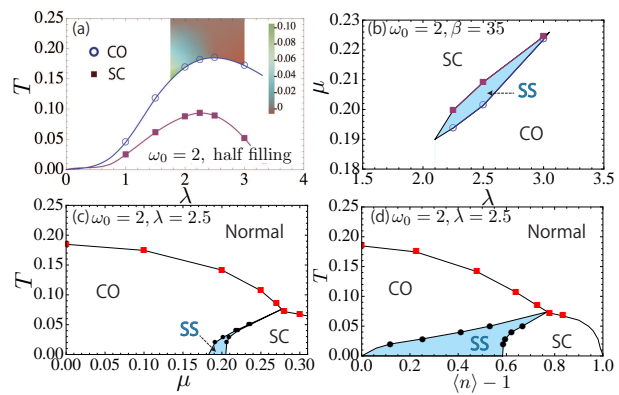}
  \caption{(Color online) For a smaller $\omega_0=2$, 
(a) Transition temperatures for CO and SC against $\lambda$ at 
half filling. Here the $T_c$ for SC is obtained by suppressing CO. In the normal state,  the dc conductivity (difference from its value at $T=0.25$ for each $\lambda$) is shown by color coding. (b) $\lambda-\mu$ phase diagram 
for the SS region for $\omega_0=2, \beta=35$. (c,d) Phase diagram of the Holstein model  plotted against (c) the chemical 
potential $\mu$, and (d) the band filling $\langle n\rangle$ for $\omega_0=2, \lambda=2.5$. Blue areas indicate the supersolid (SS) region.  }
  \label{fig:w2_sum_up}
  \end{figure}
\section{II. Dependence of phase diagram on the phonon frequency}
Here we discuss the dependence of the phase diagram on the phonon frequency $\omega_0$.
First we discuss what happens for a frequency smaller than the electron bandwidth, i.e. $\omega_0<4$. 
Figure~\ref{fig:w2_sum_up} shows phase diagrams for $\omega_0=2$. The panel (a) shows the transition temperature of CO and SC at half-filling (where we suppress CO to obtain SC).
With the frequency decreased, the positions of the peaks of the domes 
for both CO and SC 
shift to a smaller $\lambda\simeq2.5$ (see also $\omega_0=4$ data in Fig.~3 of the main text). 
The dc conductivity, displayed with color coding in Fig.~\ref{fig:w2_sum_up}(a), reveals that the metal-insulator transition occurs at some point between $\lambda=2.25$ and $\lambda=2.5$.  
Hence, the position of the metal-insulator crossover also shifts to smaller $\lambda$ and thus again coincides with the shifted peaks of the CO and SC domes. Concomitantly the SS region shifts to smaller $\lambda$, as seen in Fig.~\ref{fig:w2_sum_up}(b) 
which depicts the SS region in the plane of $\lambda$ and $\mu$ at $\omega_0=2, \beta=35$. The SS region becomes widest at $\lambda=2.5$ and  
disappears around $\lambda=2$. Hence, we conclude that as we change $\omega_0$, the SS phase, the peak of the $T_c$ dome and the metal-insulator crossover all shift in a correlated manner.

We also note that the phase diagrams for the intermediate coupling regime in the plane of temperature and chemical potential and  in the plane of temperature and density exhibit the features discussed in the main text.  
In Figs. \ref{fig:w2_sum_up}(c) and (d) we show the results for $\omega_0=2, \lambda=2.5$ as representatives  
of the phase diagrams in the intermediate coupling regime with smaller $\omega_0$. 
The SS region becomes wider with decreasing temperature, and the SC-SS phase boundary is of second order, which suggests the existence of a QCP at $T=0$. In addition, we note that the characteristic reentrant behavior can also be observed. For an even smaller $\omega_0$ such as $\omega_0=1$, we have encountered difficulties in the convergence of the self-consistency loop. Namely, the solution for consecutive iterations oscillates and does not converge within a reasonable CPU time. Although we do not have a good explanation for this phenomenon,  similar behavior has been interpreted as the tendency to ordered states with longer spatial periods, which we do not consider here \cite{Chan2009, Peters2014}.

   \begin{figure}[t]  
     \centering
  \includegraphics[width=75mm]{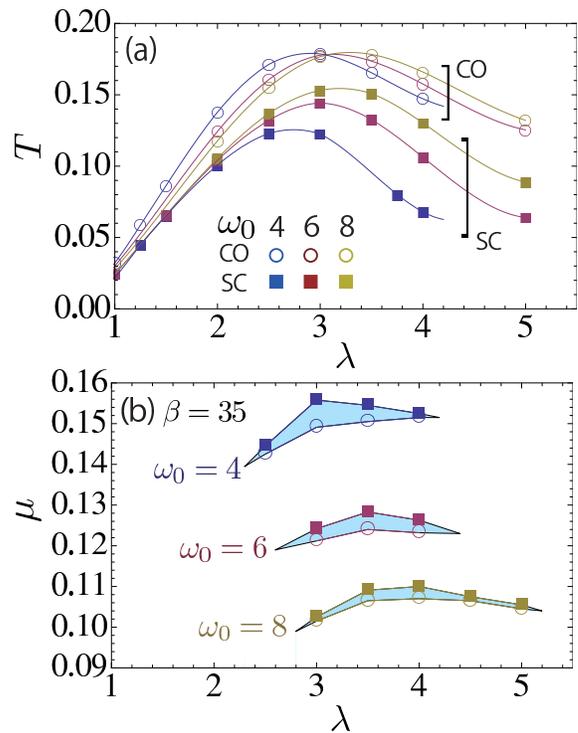}
  \caption{(Color online) (a) Transition temperatures for CO and SC against $\lambda$ for various values of $\omega_0$ at half filling. (b) $\lambda$ vs $\mu$ phase diagram 
for the SS region (shaded) at $ \beta=35$ for various values of $\omega_0$.}
  \label{fig:SS_w0dep_sum_up2}
  \end{figure}
Next we discuss the phase diagrams for larger $\omega_0$ (anti-adiabatic regime). Figure ~\ref{fig:SS_w0dep_sum_up2}(a) shows
the transition temperature of CO and SC at half-filling for various values of $\omega_0$. As the frequency increases, the position of the peak of the dome moves to larger $\lambda$. As for the position of the SS phase, we show the phase diagram in the plane of $\lambda$ and $\mu$ in  Fig.~\ref{fig:SS_w0dep_sum_up2}(b) for various values of the phonon frequency. The location of the SS phase also shifts to the larger $\lambda$ regime and stays around the position of the $T_c$ dome \footnote{The end points of the SS region (shaded area) in Fig.~3 have been established numerically, with the exception of the end point on the 
low-$\lambda$ side for $\omega_0=8$. DMFT calculations for  $\lambda=2.5, \omega_0=8$ converge very slowly. As for the other end points of the SS regime, we have checked the absence of SS at $(\lambda,\omega_0)=(5.5, 8), (4.5, 6)$ and $(2.5, 6)$.}. 
In addition to this, as $\omega_0$ increases, 
the position of the SS phase on the chemical potential ($\mu$) axis shifts to smaller $\mu$, while its width in the $\mu$ direction gradually decreases.
These results are consistent with what is expected from the attractive Hubbard model ($\omega_0\rightarrow \infty$), where the degeneracy among CO, SC and SS is lifted for a nonzero $\mu$, and doping favors SC.

\section{III. Free energy of the SS phase}

   \begin{figure}[t]  
 \centering
  \includegraphics[width=70mm]{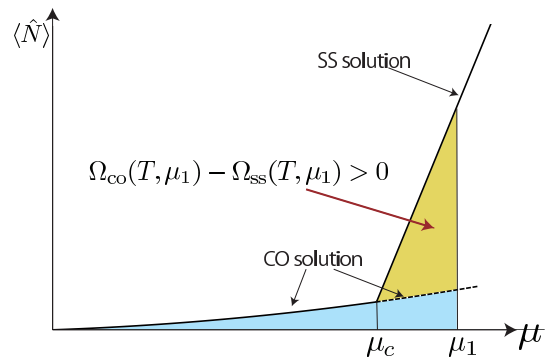}
  \caption{
  Schematic picture of the $n$ vs $\mu$ curves for the SS and CO solutions (see also Fig. 2(c) in the main text). 
  Dotted lines 
  indicates the unstable solution for CO. The yellow area represents the difference between the free energies of CO and SS.
  }
  \label{fig:free-en}
  \end{figure}

That the SS is more stable than the CO phase can be understood 
by considering the free energy 
\begin{align}
 \Omega(T,\mu)=-T \ln[\text{Tr}\exp-\beta (\hat{H}-\mu \hat{N})],
 \end{align} 
where $\hat{H}$ is the hamiltonian and $\hat{N}$ the number of particles.
 Its derivative is 
 \begin{align}
\frac{\partial \Omega(T,\mu)}{\partial \mu}=-\langle \hat{N} \rangle. \label{eq:free_der}
  \end{align}
  The situation in the vicinity of the CO-SS phase boundary is shown schematically in Fig.~\ref{fig:free-en}.
  Let us focus on $\mu=\mu_1$ in the SS region. As mentioned in the main text, one can also find a CO solution by neglecting the superconducting components, and the corresponding density is shown as a dotted line in the figure. Since the SS and CO solutions are continuously connected at the critical $\mu_c$, it follows  by integrating Eq.~(\ref{eq:free_der}) from $\mu=0$ that the yellow area corresponds to the difference between the grand canonical free energy $\Omega$ of the SS and CO solution. Since the yellow area represents a positive difference, it follows that the free energy of the SS phase is lower.

\section{IV. DMFT + IPT analysis}

   \begin{figure}[h]  
	\centering
   \includegraphics[width=85mm]{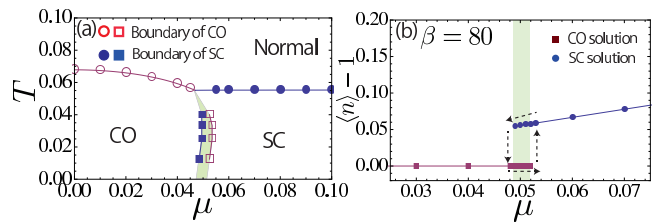}
  \caption{ (Color online)  Results of IPT + DMFT at $\omega_0=4, \lambda=1.5$. Panel (a) shows the $T$ vs $\mu$ phase diagram, and panel (b) the $\mu$-dependence of $\langle n \rangle -1$ for $\beta=80$. }
  \label{fig:w4lam3b40_ipt}
  \end{figure}
We have analyzed the weak-coupling regime using perturbative approximations in order to confirm our DMFT+CT-QMC results and to gain insights into a regime that is expensive to treat with the hybridization expansion method. Here we show, as a representative example, the results obtained by using the second-order weak-coupling expansion (IPT) as an impurity solver for DMFT, where we expand all the self-energy diagrams including the Hartree term up to fourth order in $g$ (second order in $\lambda$) \cite{Tsuji2013}. 
While the normal state of the Holstein model has been studied with 
DMFT in combination with weak-coupling approximations for the impurity 
solver \cite{Freericks1993a,Freericks1994}, we extend this analysis to ordered phases.
Figure~\ref{fig:w4lam3b40_ipt}(a) displays the IPT phase diagram and the variation of $\langle n \rangle$ with $\mu$. 
 We find that, between SC and CO, there is a first order transition with a hysteretic region  (shaded in Fig.~\ref{fig:w4lam3b40_ipt}(a)), where both CO and SC are stable DMFT solutions. Here, the hysteretic region for the two solutions has been determined as follows. For SC, we use the local Green's function for $\mu$ 
 as an initial input for $\mu-\delta\mu$ ($\delta\mu=0.001$ here). For CO, we first derive a CO solution by suppressing SC, and then add a small anomalous part ($\Phi_{SC}\simeq 0.002$) to see whether it grows or vanishes.
 In these IPT calculations, we have found no stable self-consistent SS solution -- they always converge to either SC or CO. 
We have checked that the same conclusion holds for even smaller interactions (e.g., $\lambda=1$) and confirmed the results with other perturbative schemes such as the conserving Hartree-Fock approximation \cite{Freericks1993a} and the second-order conserving approximation \cite{Freericks1994}. 

\end{document}